\documentclass[a4paper,USenglish]{article}
\usepackage{algorithm}
\usepackage{algpseudocode}
\usepackage{amsmath}
\usepackage{amsthm}
\usepackage{amssymb}
\usepackage{amsfonts} 
\usepackage{graphicx}

\usepackage{epstopdf}
\renewcommand{\Pr}{\mathop{\bf Pr\/}} 
\newcommand{\E}{\mathop{\mathbb{E}\/}}

\newcommand{\Osymbol}{{O}}
\newcommand{\BO}[1]{\Osymbol\left(#1\right)}

\newcommand{\BOx}[1]{\Osymbol(#1)}

\newcommand{\BOMx}[1]{\Omega(#1)}

\newcommand{\B}[1]{\mathcal{B}(#1)}

\newtheorem{theorem}{Theorem}
\newtheorem{lemma}[theorem]{Lemma}
\newtheorem{definition}[theorem]{Definition}
\author{Johan Sivertsen\\
  \texttt{jovt@itu.dk}}

\begin{document}
\title{Fast Nearest Neighbor Preserving Embeddings}

\maketitle

\begin{abstract}
  We show an analog to the Fast Johnson-Lindenstrauss Transform for Nearest Neighbor Preserving Embeddings in $\ell_2$.
  These are sparse, randomized embeddings that preserve the (approximate) nearest neighbors.
  The dimensionality of the embedding space is bounded not by the size of the embedded set $n$, but by its doubling dimension $\lambda$.
  For most large real-world datasets this will mean a considerably lower-dimensional embedding space than possible when preserving all distances.
  The resulting embeddings can be used with existing approximate nearest neighbor data structures to yield speed improvements.
\end{abstract}

\section{Introduction}

Many algorithmic problems become overwhelmingly difficult in high-dimensional settings.
One way of trying to combat this problem is to discover mappings that preserve the metric relevant to solving a given problem, while embedding it into a lower dimensional setting.
Most famously Johnson and Lindenstrauss~\cite{Johnson1984} showed the lemma:
\begin{lemma}[JL-Lemma\cite{Johnson1984}]
  For any integer $d>0$, and any $\epsilon  >0$, $\delta\in (0,1/2)$, for $k=\Theta(\epsilon^{-2}\log(1/\delta))$ there exists a distribution $\Pi$ such that for $k\times d$ matrices $M\sim \Pi$, for any $x\in\mathbb{R}^d$,
\[Pr\left[(1-\epsilon )\|x\|_2\leq \|Mx\|_2\leq(1+\epsilon)\|x\|_2)\right]> 1-\delta\]
\end{lemma}
The JL-Lemma shows the existence of an embedding of any set $X\subseteq \mathbb{R}^d$ into $k=\BO{\log|X|\epsilon^{-2}}$ dimensions while preserving $\ell_2$ distances up to a multiplicative $(1\pm\epsilon)$ distortion. Proofs can be found for many different $M$~\cite{FRANKL1988355,Johnson1984, Dasgupta2003,Achlioptas2003671}, including Gaussian matrices~\cite{Johnson1984, Dasgupta2003} and $\{0,\pm1\}$ matrices\cite{Achlioptas2003671}. In fact we might use any sub-gaussian distribution to fill the matrix~\cite{IN07}. 
These low dimensional embeddings can been used to speed up many fundamental high-dimensional problems like closest pair, nearest neighbor or minimum spanning tree.
They can also be used to decrease the storage requirements of a dataset when we only need to preserve norms. Further discussion and examples can be found for instance in \cite{vempala2004,Indykapplication}.
It is known that if we want to preserve the norm for all $x\in X$, the embedding dimension $k=\BOx{\log|X|\epsilon ^{-2}}$ is optimal, see \cite{DBLP:conf/icalp/LarsenN16,DBLP:journals/corr/LarsenN16,2016arXiv161000239A}.

However it might not be necessary to preserve norms for all \emph{all} points in $X$. If for example we are interested in nearest neighbor queries we require only that neighbors remain close to each other, while far away points do not get too close.
This idea was introduced and formalized as Nearest Neighbor Preserving Embeddings by  Indyk and Naor~\cite{IN07}, who also presented an embedding.
Using a full Gaussian matrix they showed that nearest neighbor distance can be preserved while embedding into fewer dimensions than in the distance preserving setting.
Specifically, $k$ is bounded by $\BO{\log \lambda_X \frac {\log(2/\epsilon)} {\epsilon^2}}$ where $\lambda_X$ is the doubling constant of $X$.
By removing the requirement that all distances be preserved we can get $k$ smaller than in the bounds discussed above~\cite{DBLP:conf/icalp/LarsenN16,DBLP:journals/corr/LarsenN16,2016arXiv161000239A}.

Another line of research has focused on improving the speed of the embeddings by using sparse matrices while keeping the distortion low\cite{DBLP:journals/jacm/KaneN14,Dasgupta:2010:SJL:1806689.1806737,Achlioptas2003671,AC09}.
Call $q<1$ the sparsity parameter.
If each entry in the used matrix is $0$ with probability $1-q$ we can improve the embedding time from $\BOx{dk}$ to expected time $\BOx{dkq}$ by sparse matrix multiplication.
A classic sparse matrix construction is the Fast Johnson Lindenstrauss Transform (FJLT) $\Phi:\mathbb{R}^d\rightarrow\mathbb{R}^k$~\cite{AC09}.
In this paper we show that the FJLT is a Nearest Neighbor Preserving embedding  with $k=\BO{\log \lambda_X \frac {\log(2/\epsilon)} {\epsilon^2}}$ and sparsity parameter $q=\BOx{\log^2{n}/d}$ for $\BO{d\log d + \epsilon^{-2}\log^3n}$ evaluation time.

\section{Preliminaries}

\begin{definition}[Nearest Neighbor Preserving Embeddings~\cite{IN07}]
  Let $\epsilon, \delta \in (0,1)$, and let $X$ be a set of points in
  $\mathbb{R}^d$. For any point $x\in X$ let $x'$ denote the point closest
  to $x$ in $X\setminus\{x\}$ under the $\ell_2$ norm.  We say that an
  embedding $\Phi:\mathbb{R}^d\rightarrow \mathbb{R}^k$ is nearest
    neighbor preserving with parameters $(\epsilon, \delta)$ if for any 
  $x \in X$, the following properties hold with probability at least
  $\delta$:
  \begin{enumerate}
  \item $\min\limits_{z\in X\setminus\{x\}}\|\Phi x-\Phi
    z\|_2\leq(1+\epsilon)\|x-x'\|_2$, and 
  \item if $\|x-y\|_2> (1+2\epsilon)\|x-x'\|_2$ for some $y \in X$ then
    $\|\Phi x-\Phi y\|>(1+\epsilon)\|x-x'\|_2$.
  \end{enumerate}
\end{definition} 

\begin{definition}[Fast Johnson-Lindenstrauss Transform~\cite{AC09}]
  \label{def:FJLT}
Let an embedding $\Phi$ be defined by a $k \times d$ matrix
  $\Phi:=PHD$ as follows: $D$ is a random $\pm1$ diagonal $d \times d$
  matrix, $H$ is the $d$-dimensional Walsh-Hadamard transform, and $P$
  is a $k \times d$ matrix populated by setting 
  \[p_{ij}=\begin{cases}X\sim\mathcal{N}(0,q^{-1}) \qquad & \text{ w.p. } q\\
    0 & \text{ w.p. } 1-q \end{cases}~~~~.\] Here the parameter $q$ is
  called the \emph{sparsity parameter} of the FJLT.
\end{definition}

\begin{definition}[Doubling constant $\lambda_X$]
  Let $B(x,r)$ denote the ball of radius $r$ centered at $x$.  The
  \emph{doubling constant} $\lambda_X$ of a point set $X \subseteq \mathbb{R}^d$
  is defined to be the smallest integer $\lambda$ such that for any $x
  \in X$, and any $r > 0$, the point set $B(x,r)\cap X$ can be covered
  by at most $\lambda$ balls $B(z,r/2)$ where $z\in X$.  We refer to
  $\log_2 \lambda_X$ as the \emph{doubling dimension} of $X$.
\end{definition}

\section{Fast Nearest Neighbor Preserving Embeddings}

Given the definitions above let us state the claim:

\begin{theorem}[Fast Nearest Neighbor Preserving Embeddings]
  \label{thm:fast-near-neighbor}
  For any $X\subseteq \mathbb{R}^d, \epsilon \in (0,1)$ where $|X|=n$ and
  $\delta\in (0,1/2)$ for some
  \[k=\BO{\frac
    {\log{(2/\epsilon)}}{\epsilon^2}\log{(1/\delta)}\log{\lambda_X}}\]
  there exists a nearest neighbor preserving embedding $\Phi:\mathbb{R}^d\rightarrow\mathbb{R}^k$ with
  parameters $(\epsilon,1-\delta)$ requiring expected \[\BO{d\log(d)+\epsilon^{-2}\log^3{n}}\] operations.
\end{theorem}

By picking $\delta$ we can fix the probability of successfully sampling an embedding that is nearest neighbor preserving and close to the expected number of operations.
Indyk and Naor presents a proof for embeddings that are constructed using
fully Gaussian matrices $G$ (see~\cite[Theorem 4.1]{IN07}). Requiring $\BOx{nd}$ operations to perform the mapping.
Our contribution will be to show how their techniques can be applied to sparse embeddings.
We first identify the properties of a map that are sufficient for the Indyk-Naor proof to hold, and then construct sparse embeddings exhibiting the properties with a bounded probability of error.

\begin{definition}
  \label{def:IN-prop}
  Let $\epsilon \in (0,1)$. We say that a distribution over maps
  $\Phi=PHD: \mathbb{R}^d \to \mathbb{R}^k$ satisfies the \emph{Indyk-Naor property}
  for a set $X \subseteq \mathbb{R}^d$ with loss $\eta \geq 0$ if with probability
  $1-\eta$ over the choice of $D$, the map satisfies that for all $x\in
  X$, $y\in X\cup \{0\}$
  \begin{enumerate}
  \item[(P1)] $\Pr_P[\|\Phi (x-y)\|_2 \not\in (1 \pm \epsilon ) \|x-y\|_2 ]\leq
    e^{-\BOMx{k\epsilon ^2}}$, and
  \item[(P2)] $\Pr_P[\|\Phi x\|_2\leq \epsilon \|x\|_2 ]\leq
    (3 \epsilon)^k$.
  \end{enumerate}
  Note that the above probabilities are taken only over the choices of
  $P$.
\end{definition}
By bounding $\eta$ with a constant we will then be able to extend the proof presented by Indyk and Naor to show the correctness of Theorem~\ref{thm:fast-near-neighbor}.
We will then need to increase $k$ by a corresponding constant to make up for the $\eta$ loss, but the order of $k$ remains unchanged.

We will show that the FJLT\cite{AC09} satisfies the Indyk-Naor properties. 
The first property to satisfy is the normal Johnson-Lindenstrauss property, but it is required to hold also for all difference vectors possible from $X$.
The second property is stronger, when $\epsilon\ll1/3$.
We will be referring to $\emph{P1}$ and $\emph{P2}$ as the Distortion and Shrinkage bound respectively.

\subsection{Smoothness}
\label{sec:smooth-setting}

We call a vector $x \in \mathbb{R}^d$ $s$-smooth if $\| x \|_\infty \leq s \, \|
x\|_2$. Note that since $H$ and $D$ are isometries
$\|HDx\|_2=\|x\|_2$.

\begin{definition}
  For any $s>0$ we say that a given diagonal matrix $D$ is in an $s$-smooth setting if
  \[\forall {x,y\in X\cup\{0\}}, \|HD(x-y)\|_\infty\leq s\|x-y\|_2.\]
\end{definition}

In this section we will bound the probability of $D$ \emph{not} being in an $s$-smooth setting for $s=\BO{\sqrt{\frac{\log n}d}}$, and then in Section~\ref{sec:distortion-bound} and \ref{sec:shrinkage-bound} we show how the Distortion and Shrinkage follow from smoothness.

Let us first consider a single vector $z=(x-y)$ where $x,y\in X\cup \{0\}$.
Assume $\|HDz\|_\infty\geq s\|z\|_2$ then there is some entry $1\leq i\leq d$ such that $|(HDz)_i|\geq s\|z\|_2$.
Let $b=\tfrac 1 {\|z\|_2}$, then $|(HDz)_{i}b|\geq s$ so we see 
\[\Pr[\|HDz\|_\infty\geq s\|z\|_2]= \Pr[\|HDzb\|_\infty\geq s]\]
where $zb$ is a unit vector. So without loss of generality we can focus on unit vectors:

\begin{lemma}
  \label{lm:inftyx}
  Given a unit vector $x$ in $\mathbb{R}^d$, 
  \[ \Pr[\| HD x \|_\infty \geq s
  ] \leq 2de^{-s^2d/2}\]
\begin{proof}
  See \cite{Ailon2009} or \cite{mitzenmacher2005probability}(p.69).
  In short, we use:
\begin{align*}
  \E[e^{sdu_1}]&=\prod_i^d\E[e^{sd h_i x_i}]=\prod_i^d\frac{1}{2}(e^{s\sqrt{d} x_i}+e^{-s\sqrt{d} x_i})\\
               &\leq\exp(s^2d\sum_{i=1}^dx_i^2/2)\\
               &=e^{s^2d\|x\|_2^2/2}
\end{align*}
In a standard Chernoff bound.
\end{proof}
\end{lemma}

As a small contribution we show a slightly better bound for our setting based on approximating the kinchine inequality constants:

\begin{lemma}
  \label{lm:kinchine}
  Given a unit vector $x$ in $\mathbb{R}^d$, 
  \[ \Pr[\| HD x \|_\infty \geq s
    ] \leq de^{-s^2d/2.2}\]
  For $s>3/\sqrt{d}$.
\begin{proof}
  Let $u=HDx=(u_1,..,u_d)^T$, so $u_1=\sum_i^dh_ix_i$ where the $h_i$ are i.i.d $\pm d^{-1/2}$.
  For all $p\geq1$:
  \begin{equation}
    \label{eq:1}
    \Pr[|u_1| \geq s]
    = \Pr\left[\left|\sum_i^d\pm x_i\right|^p \geq (\sqrt{d}s)^p\right]
    \leq \frac{\E\left[\left|\sum_i^d\pm x_i\right|^p\right]}{(\sqrt{d}s)^p}
  \end{equation}

  By the kinchine inequality, for $p\geq9$:

  \[\E\left[\left|\sum_i^d\pm x_i\right|^p\right]\leq B_p\|x\|_2^p\leq\left(\sqrt{p/2.5}\|x\|_2\right)^p\]

  Since $B_p=\max\{1,2^{p-2/2}\frac{\Gamma(\frac{p+1}{2})}{\Gamma(3/2)}\}$~\cite{Nazarov2000,Haagerup1981} which is $\leq\sqrt{p/2.5}^p$ for $p\geq 9$.

  Hence:

  \[\Pr[|u_1| \geq s]\leq\left(\frac{p}{2.5ds^2}\right)^{p/2}\]
  By setting $p=\max\{9,s^2d)\}$ we get:
  \[\Pr[|u_1| \geq s]\leq e^{-{s^2d}/2.2}\]

  for $s\geq3/\sqrt{d}$.  Union bound over $u$ gives the claim.
  
\end{proof}

\end{lemma}

\subsection{Fixing $s$ and $q$}

Now let $s=d^{-1/2}\sqrt{c\ln(n^2d)}$, then by Lemma~\ref{lm:kinchine} for $x\neq y$:

\[\Pr[\exists x,y \in X\cup \boldsymbol{0},\|HD(x-y)\|_\infty\geq s\|x-y\|_2]\leq \frac{n^2d}{e^{c\ln(n^2d)/2.2}}=e^{-c/2.2}\]

If we fix $s$ by setting $c=7$ we observe that $D$ is in an $s$-smooth setting with probability at least $\frac {19} {20}$.
Recall that we used $s\geq 3/\sqrt{d}$ which holds in our setting as long as $n^2d\geq3.7$.
In the following we will let $\Phi$ be a FJLT embedding constructed by setting $q=\min\left(c's^2,1\right)$ where $c'>0$ is some universal constant.
We will then show that if $D$ is $s$-smooth, this setting of $q$ makes $\Phi=PDH $ satisfy the distortion and shrinkage bounds.

\subsection{Distortion bound}
\label{sec:distortion-bound}

\begin{lemma}[Distortion bound] 
  For any $x,y\in X\cup \{0\}$ if $D$ is in an $s$-smooth setting, for $\epsilon>0$:
  \label{lm:distortion}
  \[\Pr\left[|\|\Phi(x-y)\|_2\notin (1\pm\epsilon)\|x-y\|_2 \right]\leq e^{-\BOMx{k\epsilon ^2}}\]

  \begin{proof}
    The distortion bound is the main result in~\cite{AC09}.
    
\end{proof}
\end{lemma}

\subsection{Shrinkage bound}
\label{sec:shrinkage-bound}

The shrinkage bound is stronger than the distortion bound for large $\epsilon$.
We will need it later to confine the probability of any of an infinite series of events happening to a small constant.

\begin{lemma}[Shrinkage bound]
  For a fixed vector $x\in X$, if $D$ is in an $s$-smooth setting,
  for $\epsilon \in (0,1)$:
  \label{lm:fractionshrink}
  \[\Pr_P\left[\|PHD x\|_2\leq \epsilon \|x\|_2 \right]\leq \left({3}{\epsilon }\right)^k\]
\end{lemma}
By the regular scaling of Gaussian with their standard deviation (See Lemma~\ref{lm:generalizednonstd}), it is clear that for an upper bound on:
\[\Pr[\sum_i^ky_i^2\leq t]\]  
where the $y_i\sim\mathcal{N}(0,Z_i/q)$ we only need to lower bound the $Z_i$.
I.e.
\begin{lemma}
  \label{lm:non-to-std}
If $D$ is in an $s$-smooth setting and $Z_i\geq q/2$ for all $i\in [k]$, then $\forall t\geq 0$:
\[\Pr[\|PHDx\|_2^2\leq t]\leq\Pr[\|Gx\|^2\leq 2t]\]

\begin{proof}
For $i\in\{1,\cdots,k\}$ assume $Z_i\geq \frac q2$ and let $X_i\sim\mathcal{N}(0,1)$, then:
\begin{align*}
  \Pr\left[\|PHDx\|_2^2\leq t\right]&=\Pr\left[\sum_{i=1}^ky_i^2\leq t\right]\\
              &\leq \Pr\left[\sum\left(\sqrt{\dfrac 1 2}X_i\right)^2\leq t\right]=\Pr\left[\sum X_i^2\leq 2t\right].
\end{align*}
Where the first equality follows from $D$ being in an $s$-smooth setting and the inequality from the bound on the $Z_i$.
\end{proof}
\end{lemma}

For our setting of $q$, $\Pr\left[ \forall i \in [k], Z_i\geq q/2 \right] \geq \frac{19}{20}$(Lemma 3 of~\cite{Ailon2009}).
If we combine this bound with Lemma~\ref{lm:non-to-std} we are ready to prove Lemma~\ref{lm:fractionshrink}.

\begin{proof}
Let $z\in X$ and let $x=z\|z\|_2^{-1}$.
    \begin{align*}
      &\Pr\left[\|PHDz\|_2\leq \epsilon \|z\|_2\right]=\Pr\left[\|PHDx\|_2\leq {\epsilon }\right] =\\
      &\Pr\left[\|PHDx\|_2^2\leq {\epsilon ^2}\right] \leq \Pr\left[\sum_i^kX_i^2\leq 2 {\epsilon ^2}\right] \tag{by
        Lemma~\ref{lm:non-to-std} }
    \end{align*}
    Where $X_i\sim\mathcal{N}(0,1)$. In general for $s,t>0$ we know:

    \begin{align*}
      \Pr\left[\sum_i^kX_i^2\leq t\right]&=\Pr\left[e^{-s\sum X_i^2} \geq e^{-st}\right]\leq \frac {\E\left[e^{-s\sum X_i^2}\right]}{e^{-st}}\\
                                         &= e^{st}\prod_{i=1}^k\E\left[e^{-sX_i^2}\right]=e^{st}(1+2s)^{-k/2}
    \end{align*}

    Where the last step uses that $\E\left[e^{-sX_i^2}\right]=\frac 1 {\sqrt{1+2s}}$ for $-1/2 \leq s \leq \infty$.(See ~\cite{Dasgupta2003})

    Now to minimize we differentiate w.r.t s:

    \begin{align*}
      &t e^{st}(1+2s)^{-\frac{k}{2}}+2(-\frac k2)e^{st}(1+2s)^{-\frac k2 -1}=0 \Leftrightarrow t=k(1+2s)^{-1}\\
      &\Rightarrow s=(k/t-1)/2
    \end{align*}

    So $e^{st}(1+2s)^{-k/2} =   e^{(k-t)/2}(k/t)^{-k/2}= e^{-t/2}(\frac k {et})^{-k/2} \leq(et)^{k/2}$.
    Now plug in $t=2\epsilon^2$ and we have
   \[\Pr\left[\sum_i^kX_i^2\leq 2\epsilon^2\right]\leq (2e\epsilon^2)^{k/2}\leq(3\epsilon )^k\]
    
\end{proof}

\subsection{Embedding properties}

We have now seen how the Distortion and Shrinkage bounds follow from two events:

First $D$ must be in an $s$-smooth setting. Secondly all $Z_i$ must be within a constant factor of $q$.
By Lemma~\ref{lm:kinchine} the first event happens with probability at least $19/20$ when setting $s=\sqrt{7\frac{\lg(n^2d)}{d}}$, assuming $n^2d\ge 3.7$.
By choosing $q$ corresponding to $s$ as in\cite{Ailon2009}, the second event occurs with probability at least $19/20$(See Lemma 3 of~\cite{Ailon2009}).
For the chosen parameters $\Phi=PHD$ satisfies the Indyk-Naor properties with probability $\left(\frac{19}{20}\right)^2>9/10$.

We can then move on to prove Theorem~\ref{thm:fast-near-neighbor} by showing:

\begin{theorem}[Fast Nearest Neighbor Preserving Embeddings]
  \label{thm:fast-near-neighbor-by-FJLT}
  For any $X\subseteq \mathbb{R}^d, \epsilon \in (0,1)$ there exists
  \[k=\BO{\frac
      {\log{(2/\epsilon)}}{\epsilon^2}\log{(1/\delta)}\log{\lambda_X}}\]
  such that for every $x\in X $ if $x'$ denotes the point closest
  to $x$ in $X\setminus\{x\}$ under the $\ell_2$ norm. with probability at least $\delta$:
    \begin{enumerate}
  \item $\min\limits_{z\in X\setminus\{x\}}\|\Phi x-\Phi z\|_2\leq(1+\epsilon)\|x-x'\|_2$, and 
  \item if $\|x-y\|_2> (1+2\epsilon)\|x-x'\|_2$ for some $y \in X$ then $\|\Phi x-\Phi y\|>(1+\epsilon)\|x-x'\|_2$.
  \end{enumerate}
  Where $\Phi=PHD$ is a FJLT matrix, with expected \[\BO{d\log(d)+\epsilon^{-2}\log^3(n)\log(2/\epsilon)}\] embedding time.

\begin{figure} 
 \includegraphics[width=0.3\paperwidth]{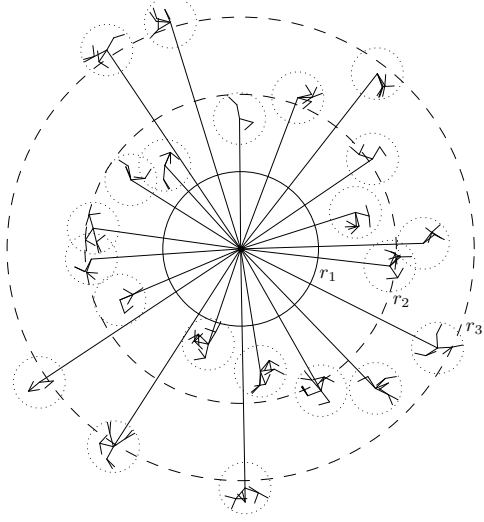}
  \caption{An illustration of the spanning tree construction used in the proof of Theorem~\ref{thm:fast-near-neighbor-by-FJLT}.}
    \label{fig:spantree}
\end{figure}

\begin{proof}
  Let $\Phi=PHD$ follow Definition~\ref{def:FJLT} with $q=\min\{\BOx{d^{-1}\ln(n^2d)},1\}$ so $\Phi$ satisfies the Indyk-Naor properties as pr. Definiton~\ref{def:IN-prop} with probability at least $9/10$.
  The proof then follows from~\cite[Theorem 4.1]{IN07}.
  For completeness we include an extended version of the proof here.
  For familiar readers, the only difference in this version is in making the spanning tree construction explicit.

  Without loss of generality let $x=0$ and $\|x'\|_2=1$.
  To show the first property let $y\in X$ satisfy $\|y\|_2=1$, then by the distortion bound, $\Pr[\|\Phi y\|\geq(1+\epsilon)]\leq e^{-\BOMx{k\epsilon^2}}$.
  So for some universal $C$, setting $k\geq C\ln(1/\delta)/\epsilon ^2$ we get:
  \[\Pr[\min\limits_{z\in X\setminus\{x\}}\|\Phi x-\Phi z\|_2>(1+\epsilon)\|x-x'\|_2]<\delta/2\]

  To show the second property we construct a spanning tree of $\left(X\setminus \B{x,1+2\epsilon }\right)\cup \{0\}$ with $0$ at the root.
  Let $r_i=1+(i+2)\epsilon$. Consider the annuli:
  \[ X_i=X\cap\B{0,r_{i+1}}\setminus\B{0,r_{i}},\text{ for } i\geq1 \]
  By the definition of $\lambda_X$, for any $i$ we can construct a minimal set $S_i\subseteq X$ such that $X_i\subseteq \cup_{t\in S_i}\B{t,\epsilon/4}$ and $|S_i|\leq \log_2(\frac {4r_i} \epsilon)$.
  The first level of the tree consists of an edge between $x$ and each $t\in S_i$ for all $i\geq0$.
  From each $t$ a spanning tree is build on the points in $\B{t,\epsilon/4}$ with $t$ at the root, as described in lemma~\ref{lem:embedding-properties}.
  Figure~\ref{fig:spantree} illustrates the construction.
  Some ordering is imposed on the $t$ points so points in overlapping balls are only spanned once.

  We can then restate the second property as $\exists i\geq0,\exists x\in X_i . \|\Phi x\|\leq 1+\epsilon $, at least one of two events took place:
  \begin{enumerate}
  \item $\exists i\geq0,\exists t\in S_i. \|\Phi t\|_2\leq 1+\epsilon+\frac \epsilon 4 (1+\sqrt{i})$
  \item $\exists i\geq0,\exists t\in S_i, \exists x \in \B{t,\frac \epsilon 4}\cap X. \Phi x \notin \B{\Phi t,(1+\sqrt{i})\frac \epsilon 4}$
  \end{enumerate}
  Since $\|t\|_2\geq r_i - \frac \epsilon 4$ there is some constant $C$ such that:
  \[
    \frac {\|\Phi t\|}{\|t\|_2} = \frac {1 + (1+\sqrt{i})\epsilon/4  + \epsilon } { 1+ (2+i)\epsilon - \epsilon /4} \leq
    \begin{cases}
      1-\epsilon/8 \text{ for } i\leq 1/\epsilon^2\\
      C/\sqrt{i} \text{ for } i > 1/\epsilon^2
  \end{cases}
\]

Fix some $i$. Using the distortion and shrinkage bounds:
\begin{align*}
  &\Pr\left[\exists t\in S_i, \|\Phi t\|_2\leq1+ (1+\sqrt{i})\epsilon/4 + \epsilon\right]\\
           &\leq\begin{cases}
             \lambda_x^{\log_2(4r_i/\epsilon)}e^{-ck\epsilon^2}\text{ for } i\leq 1/\epsilon ^2\\
             \lambda_x^{\log_2(4r_i/\epsilon)}(3C/\sqrt{i})^k\text{ for } i> 1/\epsilon ^2
           \end{cases}\\
             &\leq\begin{cases}
             e^{-c'k\epsilon ^2}\text{ for } i\leq 1/\epsilon ^2\\
             i^{-c'k} \text{ for } i> 1/\epsilon ^2
           \end{cases}\\
\end{align*}

For $k\geq \frac {c''} {\epsilon^2}\log(2/\epsilon)\log(\lambda_X)$ where $c''$ is some universal constant.
For the second event lemma~\ref{lem:embedding-properties} gives:
\[\Pr[\exists t, \exists y \in \B{t,\frac \epsilon 4}\cap X, \Phi y \notin \B{\Phi t,(1+\sqrt{i})\frac \epsilon 4}]\leq\lambda_x^{\log_2(4r_i/4)}e^{-ck(1+i)}\leq e^{-c'k(1+i)}\]
So there is some $c'''$ where the first event is most likely. Hence:
\[
\Pr[\exists x\in X_i. \|\Phi x\|_2\leq 1+ \epsilon ]\leq  
\begin{cases}
  2e^{-c'''k\epsilon ^2}\text{ for } i\leq 1/\epsilon ^2\\
  2i^{-c'''k} \text{ for } i> 1/\epsilon ^2
\end{cases}\\
\]

Summing over all the $i$ we get:
\begin{align*}
  \Pr[\exists i&\geq 0, \exists x \in X_i . \|\Phi X\|_2 \leq 1+\epsilon ] = \sum_i^{\infty} \Pr[\exists x\in X_i. \|\Phi x\|_2\leq 1+ \epsilon ]\\
         &\leq \frac 2 {\epsilon^2} e^{-c'''k\epsilon ^2} + \sum_{i>1/{\epsilon ^2}}  2i^{-c'''k}\leq \delta/2
\end{align*}

for some $k\geq \log(1/\delta)\frac {\tilde{c}} {\epsilon^2}\log(2/\epsilon)\log(\lambda_X)$ where $\tilde{c}$ is some large enough constant.
The number of operations required for embedding $x$ is $\BOx{d}$ for the diagonal matrix $D$, $\BOx{d\log{d}}$ for $H$ using the Walsh-Hadamard transform~\cite{Fino:1976:UMT:1311952.1312575} and finally $\BO{|P|}$ where $|P|$ is the number of non-zero entries.
By our setting of $q$, $|P|\sim\mathrm{Bin}(kd,q)$  so
\begin{align*}
  \E\left[|P|\right]&=kdq\\
                    &=\BOx{\epsilon^{-2}\log(\lambda_X)\log(2/\epsilon)\log^2(n)}\\
                    &=\BOx{\epsilon^{-2}\log^3(n)\log(2/epsilon)}
\end{align*}
\end{proof}
\end{theorem}

\begin{lemma}
  \label{lem:embedding-properties}
  Let $X$ be a subset of the unit ball in $\mathbb{R}^d$, including $0$.
  Then there exists universal constants $c,C>0$ such that for $\epsilon>0$ and $k\geq C\log{\lambda_X}$:
  \[
    \Pr[\exists x\in X, \|PHDx\|_2\geq(1+\epsilon)]\leq e^{-ck(1+\epsilon)^2}.
  \]
  \begin{proof}
    The proof is given in ~\cite[Lemma 4.2]{IN07}.
    We include a spanning tree version here for completeness.
    We build a spanning tree $T$ on $X$ with root $0$ in the following way:
    Define sets for each possible level of the tree, $L_0,L_1,\ldots\subseteq X$.
    Let $L_0={0}$.
    To build $L_{j+1}$, for every point $t\in L_j$ let $S_{t}$ be the minimal size set such that $\cup_{s\in S_{t}}B(t,2^{-j-1})\cap X$ covers all of $B(t,2^{-j})\cap X$.
    By the definition of doubling constant we know that $|S_{t}|\leq \lambda_X$.
    Connect $t$ to every point in $S_t$, if some $S_t$ sets overlap only a single connection is made to avoid cycles.
    Let $L_{j+1} = \cup_{t\in L_j}S_t$. We observe that $0<|L_j|\leq \lambda_X^j$.

    Now let $E(T)$ denote the edges in $T$.
    Let $E_j$ be the subset of $E(T)$ with one node in $L_{j}$ and the other in $L_{j+1}$, by the construction of the tree $\forall e\in E_j$ we have $\|e\|_2\leq 2^{-j+1}$.
    For every $x\in X$ denote the unique path from $0$ to $x$ in $T$ by $p(x)\subseteq E(T)$.
    For $0\leq j\leq |p(x)|$ let $p_j(x)\in L_j$ be the vertex on the path at level $j$, for $j>|p(x)|$ let $p_j(x)=x$.
    We can then compose $x$ as $\sum_{j=0}^{\infty}\left(p_{j+1}(x)-p_{j}(x)\right)$, the first $|p(x)|$ steps corresponding to edges in $E(T)$, and the remaining steps having $0$ contribution.
    The argument then follows~\cite{IN07}:
    \begin{align*}
      \Pr[\exists x\in& X, \|PHDx\|_2\geq(1+\epsilon)]\\
              &\leq\Pr\left[\exists x \in X, \exists j\geq 0, \|PHD(p_{j+1}(x)-p_{j}(x))\|_2\geq\frac {(1+\epsilon)} 3 \left(\frac 3 2 \right)^{-j}\right] \\
              &=\sum_{j=0}^\infty \Pr\left[\exists e\in E_j, \|PHDe\|_2\geq \frac {1+\epsilon} 3 \left(\frac 3 2 \right)^{-j}\right]\\
              &\leq\sum_{j=0}^\infty \Pr\left[\exists e\in E_j, \|PHDe\|_2\geq \frac {1+\epsilon}6 \left( \frac 43 \right)^j\|e\|_2\right]\\
              &\leq\sum_{j=0}^\infty \lambda_X^{2j}\Pr\left[\|PHDx\|_2\geq1+\frac {1+\epsilon}6 \left( \frac 43 \right)^j-1\right]\text{,for any unit vector $x$}\\
              &\leq\sum_{j=0}^\infty \lambda_X^{2j}e^{-ck(1+\epsilon)^2(4/3)^{2j}/100}\leq e^{-ck(1+\epsilon)^2}
    \end{align*}
    For $k\geq C\log\lambda_X+1$.
    Crucially the second last step uses that$|E(T)|=|X|-1$.
    We can then use Lemma~\ref{lm:kinchine} to see that $D$ is in a smooth setting with constant probability, for our setting of $s$ at least $\frac {19}{20}$.
    The last step then follows from Lemma~\ref{lm:distortion}.
  \end{proof}
  
\end{lemma}

\section{Conclusion and future work}

In this paper we present embeddings that combine the low dimensional embedding space achieved by Nearest Neighbor Preserving Embeddings~\cite{IN07} with a speedup of the embedding runtime achieved by a Fast-JL construction~\cite{AC09}.
This results in embeddings that are faster than fully Gaussian Nearest Neighbor Preserving Embeddings and use fewer dimensions than any Johnson-Lindenstrauss type embedding.

The benefit of Nearest Neighbor Preserving Embeddings generally depends on the difference between $n=|X|$ and $\lambda_X$.
While $\lambda_X$ is always upper bounded by $n$ it can often be much smaller, this helps to explain why some datasets can be successfully embedded into much fewer dimensions, and much faster, than theoretical results looking only on $|X|$ can explain.
For datasets with low doubling dimension we can expect to find fast embeddings into a low number of dimensions, even if the dataset is very large.

While the number of rows in the embedding matrix is independent of $n$, the sparsity of the matrix is not.
This happens because we must ensure that all $\BOx{n^2}$ possible edges in the constructed spanning trees used in lemma~\ref{lem:embedding-properties} are smooth.
Future work could focus on alternative constructions using closer to $\BOx{n}$ edges or other ways of increasing the sparsity.

\subsubsection*{Acknowledgements}
The author would like to thank Anupam Gupta for much fruitful discussion during the writing of this paper.

\bibliographystyle{plain}
\bibliography{biblio}

\section{Appendix: Properties of Gaussians}

\begin{lemma}
  \label{lm:nonstdgauss}
  Let $X\sim \mathcal{N}(0,x)$ and $Y\sim \mathcal{N}(0,y)$.
  Then $\forall t>0$:
  \begin{align}
    y\geq x &\Rightarrow \Pr[X^2\leq t]\ge \Pr[Y^2\leq t]\\
    y\leq x &\Rightarrow \Pr[X^2\leq t]\leq \Pr[y^2\leq t]
  \end{align}
With equality exactly when $x=y$.  

  \begin{proof}
    Let $y\geq x$:
    \begin{align}
      \Pr[X^2\leq t]=\Pr[X\leq\sqrt{t}]-\Pr[X\leq-\sqrt t]&\geq \\
      \Pr[Y\leq\sqrt t]-\Pr[X\leq-\sqrt t]&\geq \\
      \Pr[Y\leq\sqrt t]-\Pr[Y\leq-\sqrt t]&=\Pr[Y^2\leq t]
    \end{align}
    Similarly in the other direction when $y\leq x$.
    \end{proof}
\end{lemma}
    We can generalize to sums of such variables:
    
    \begin{lemma}
      \label{lm:generalizednonstd}
      For any integer $k\geq 1$. Let $X=\sum_{i=1}^k X_i^2$ where $X_i\sim\mathcal{N}(0,x_i)$ and $Y=\sum_{i=1}^kY_i^2$ where $Y_i\sim\mathcal{N}(0,y_i)$.
      Then if $y_i\geq x_i$ for all $i\in\{1,\cdots,k\}$ we have: 
      \[\Pr[Y\leq t]\leq\Pr[X\leq t].\]
      \begin{proof}
        We show a standard proof by induction.
        Define a new variable $S_l=\sum_{i=1}^{k-l}X_i^2+\sum_{j=k-l+1}^kY_{j}^2$.
        As a base case set $l=1$:
        \[\Pr[S_1\leq t]= \Pr[Y_k^2+\sum_{i=1}^{k-1}X_i^2\leq t]\leq\Pr[S_0\leq t].\]
        By fixing $X^2_i$ for $1\leq i\leq k-1$ and using lemma.~\ref{lm:nonstdgauss}.
        And generally for all integers $l>0$ up to $l=k$:

        \[\Pr[S_l\leq t]=\Pr[S_{l-1}\leq t ]\]
        By fixing everything but the $l$'th variable and using lemma.~\ref{lm:nonstdgauss}.
        We arrive at $\Pr[Y\leq t]=\Pr[S_k\leq t]\leq\Pr[S_0\leq t]=\Pr[X\leq t]$.        
      \end{proof}
    
    \end{lemma}

    \begin{lemma}[2-stable Gaussians]
      \label{lm:2stable}
Let $X,X_1,..,X_d\sim \mathcal{N}(0,\sigma^{2})$ and $u\in \mathbb{R}^d$, then:
\[\sum_{i=1}^d u_iX_i\sim \|u\|_2X \sim \mathcal{N}(0, \sigma^2\|u\|_2^2)\]
\begin{proof}
See \cite{zolotarev1986one}.
\end{proof}

\end{lemma}

\newpage

\end{document}